\documentclass[10pt,conference]{IEEEtran} 
\IEEEoverridecommandlockouts

\usepackage[pdftex]{hyperref}
\makeatletter
\AtBeginDocument{
  \hypersetup{
    pdftitle = {Streaming GST with the extended Kalman filter},
    pdfauthor = {Marceaux and Young}
  }
}
\makeatother
\usepackage{amsfonts,amsmath,amssymb,amsthm}
\usepackage{array,bm,color}
\usepackage{epsfig,graphicx,nomencl,revsymb4-1,upgreek,url}
\usepackage{algorithm}
\usepackage{algpseudocode}
\usepackage{bbm}
\usepackage{subcaption}
\usepackage[style=ieee]{biblatex}

\graphicspath{{./figures/}}
\hypersetup{colorlinks=true, pdfauthor=J. P. Marceaux and Kevin Young, pdftitle=TITLE}

\newcommand{\sbra}[1]{\ensuremath{\left\langle\left\langle{#1}\right\vert\right.}}

\newcommand{\sket}[1]{\ensuremath{\left.\left\vert{#1}\right\rangle\right\rangle}}

\renewcommand{\cdot}{}

\def\BibTeX{{\rm B\kern-.05em{\sc i\kern-.025em b}\kern-.08em
    T\kern-.1667em\lower.7ex\hbox{E}\kern-.125emX}}

\begin{document}


\title{Streaming quantum gate set tomography using the extended Kalman filter}

\author{\IEEEauthorblockN{J. P. Marceaux}
\IEEEauthorblockA{\textit{Berkeley Center for Quantum Information and Computation} \\ 
University of California, Berkeley,\\
Berkeley, California 94720, USA \\
j.p.marceaux@berkeley.edu }
\and
\IEEEauthorblockN{Kevin Young}
\IEEEauthorblockA{\textit{Quantum Performance Laboratory} \\
Sandia National Laboratories \\
Livermore, CA 94550, USA\\
kyoung@sandia.gov}
}
\maketitle

\thispagestyle{plain}
\pagestyle{plain}


\begin{abstract}
\noindent
Closed-loop control algorithms for real-time calibration of quantum processors require efficient filters that can estimate physical error parameters based on streams of measured quantum circuit outcomes. Development of such filters is complicated by the highly nonlinear relationship relationship between observed circuit outcomes and the magnitudes of elementary errors. In this work, we apply the extended Kalman filter to data from quantum gate set tomography to provide a streaming estimator of the both the system error model and its uncertainties. Our numerical examples indicate extended Kalman filtering can achieve similar performance to maximum likelihood estimation, but with dramatically lower computational cost. With our method, a standard laptop can process one- and two-qubit circuit outcomes and update gate set error model at rates comparable with current experimental execution. 
\end{abstract}

\begin{IEEEkeywords}
Quantum tomography, Kalman filtering, quantum calibration, QCVV
\end{IEEEkeywords}



\section{Introduction}
\label{sec:introduction}
\begin{figure}[H]
    \centering
    \includegraphics[width=.9\columnwidth]{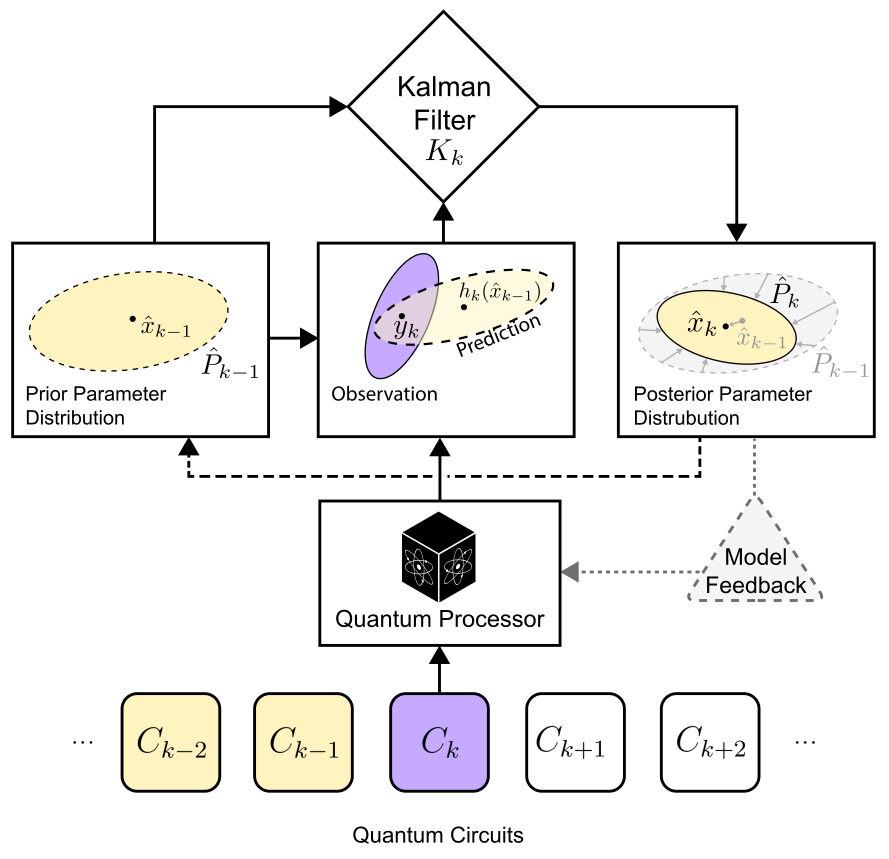}
    \caption{Closed-loop control techniques benefit from online filters that can update model parameters in real time as data is collected. This work adapts the extended Kalman filter for the purpose of streaming estimation of error rates (and their uncertainties) in gate-model quantum processors. This approach offers an alternative to the batched maximum likelihood estimation utilized in, eg. gate set tomography.}
    \label{fig:ClosedLoopControl}
\end{figure}

\noindent 
Efficient, closed-loop stabilization protocols that utilize active experimental feedback will be necessary for future quantum processors to enable rapid calibration and to maintain error rates below the threshold for fault tolerance. A key element of closed-loop control is a filter that can estimate model parameters from noisy data in a streaming fashion, conventionally reffered to as an ``online estimator.'' However, standard approaches for estimating error rates in quantum computers use tomographic estimation techniques that rely on post-processing large amounts of batched data to obtain reliable estimates \cite{Nielsen2021-el}. Recursive filters, such as the Kalman filter \cite{Kalman1960-jj}, offer a compelling alternative and have a long history of performing the streaming parameter estimation that underlies many industrial control techniques. In this work, we show that quantum gate set tomography (GST) \cite{Nielsen2021-el} can be efficiently performed using an extended Kalman filter \cite{Daum2021-wi} to simultaneously estimate error rates in quantum operations and their uncertainty (see Fig.~\ref{fig:ClosedLoopControl}).  

GST is a quantum characterization technique that allows for precise estimation of the rates of elementary errors suffered by a quantum processor. Unlike randomized characterization techniques, which can only reliably estimate stochastic noise, GST can provide high-precision estimates of the coherent errors that often arise from drift and miscalibration, such as detunings and over-rotations. GST achieves this precision by measuring outcome distributions of different quantum circuits composed of short sequences of gates that are repeated many times. Data collected from running these circuits is then used to fit the parameters of an error model, most commonly with batched maximum likelihood estimation (MLE). Uncertainty in the resulting estimate can be computed by examining the shape of the likelihood function around the MLE. Our method replaces this batched MLE with an online, recursive filter that updates the estimate and estimated uncertainty after each circuit that is run, and whose performance in simulation rivals the fits provided by MLE. 

Bayesian inference \cite{Bertsekas2008-rl} is a viable alternative to maximum likelihood estimation that naturally incorporates new information in a streaming fashion. However, the full Bayesian inference problem without simplification is computationally complex and ill-suited for real-time characterization within the short time-scales of quantum gate operation. However, simplifying model assumptions can greatly reduce the complexity of Bayesian inference \cite{Evans2022-jz}. In the case of the extended Kalman filter, the assumptions of a) a linearized model, b) Gaussian noise, and c) a Gaussian prior reduce computation and optimization of the Bayesian posterior to a series of inexpensive matrix operations \cite{Meinhold1983-oj}. Though the general gate set tomography problem formally violates the assumption of linearity and Gaussian noise, we detail several approximations and experiment design choices that can successfully embed GST within the framework of Kalman filtering. 

We propose Kalman filtering for quantum device characterization as an online protocol that can be implemented on classical control hardware running concurrently with quantum circuit execution. In Section \ref{sec:background}, we review the basics of gate set tomography and Kalman filtering necessary for this work. In Section \ref{sec:kf_gst}, we develop a Kalman filter implementation of GST and emphasize the necessary approximations. We present numerical results in Section \ref{sec:results} that demonstrate extended filtering can perform comparably to MLE. We finally sumarize several extensions and alternative approaches to our method that may be useful in deploying streaming gate set tomography on real devices. 

\subsection{Related work}
The study of online estimation using Kalman filters and related techniques is a well-developed discipline with numerous introductory textbooks \cite{Grewal2014-gs, The_Analytic_Sciences_Corporation1974-zy}. Since its original formulation \cite{Kalman1960-jj} in 1960, the Kalman filter has inspired many reformulations including sigma point (unscented) \cite{Julier1997-mu}, ensemble \cite{Evensen2003-lr}, invariant \cite{Barrau2018-jo} Kalman-type filters, and $H_\infty$ filters \cite{Simon2006-dd}. In this work, we derive an extended Kalman filter for the gate set tomography estimation problem, which may be seen as a first step towards more sophisticated techniques. 

Online estimation techniques have already been extensively explored in the context of quantum state and process tomography \cite{Blume-Kohout2010-ax, Granade2016-gq}. Some attention has been paid to quantum Kalman filters in the context of continuous weak quantum measurement \cite{Emzir2017-uo, Iida2010-eo, Agarwal2019-jf, Geremia2003-rn}. In the context of discrete projective measurements, the regime considered in this work, the Kalman filter has been used to estimate error bars in quantum state tomography \cite{Audenaert2009-dr}. More generally, Ref. \cite{Gupta2021-kw} showed that much of the classical theory of nonlinear filters is directly applicable to the estimation problems that arise frequently in quantum computing. To our knowledge, no work has considered Kalman filtering in the context of gate set tomography. 

Online approaches for estimating errors in quantum gate sets have been explored, but to a lesser degree than in state tomography. Ref. \cite{Di_Matteo2020-in} developed a particle filter for online GST estimation, but due to the computational complexity of particle filters, it is unlikely such an approach would be feasible for real-time characterization. More recent work \cite{Evans2022-jz}, demonstrated a Fast Bayesian Tomography (FBT) algorithm capable of real-time characterization of quantum gate sets with a Bayesian inference technique based on a simplified Gaussian model, which is similar to the method we develop here. However, the method we develop in this work is independent and explicitly built on the Kalman filter, which provides a robust theoretical foundation. There are also some further technical details in the FBT algorithm, such as the use of ``linearization noise'' and sampling in addition to Jacobian calculations that may make FBT sub-optimal in performance and resource consumption. 

An essential ingredient in our online protocol is model linearization, where a nonlinear model is approximated by its first order Taylor expansion. Ref. \cite{Gu2021-zx} utilized random circuits and linearized about the target (ideal) model to develop a fast estimation algorithm for errors in quantum gates. Their work used a ``design matrix'' or the matrix of the first derivative of model probabilities with respect to parameters, i.e., the Jacobian. A similar object appears frequently here. 

\section{Introduction to GST and the Kalman filter}
\label{sec:background}

In this section we provide brief introductions to both gate set tomography (GST) and the Kalman filter. We cover only the material necessary to show how Kalman filters can be applied to GST. For more extensive reviews of GST, see \cite{Nielsen2021-el}. For Kalman filtering, see \cite{Simon2006-dd}. We summarize our notation in Table \ref{tab:notation} below.

\begin{table}
    \centering
    \begin{tabular}{c|l}
         Symbol & Description \\
         \hline
         $|\rho \rangle \rangle$ & State preparation \\
         $G_i$ & Process matrix representation of a gate \\
         $\langle E_i|$  & Measurement effect \\
         $C_k$ & Quantum circuit (a product of gates) \\
         $x$ & Vector of gate set error parameters \\
         $s^{(k)}_j$ &  $j$th categorical random variable sampled from circuit $k$\\ 
         $y_k$ & Observation -- average frequency of measurement outcomes \\
         $h_k$ & Observation model -- maps error parameters to predictions \\
         $\mathcal{Y}_k$ & History of observations, $y_1, y_2, ..., y_k$ \\
         $\hat{x}_k$ & Estimate of error parameters given previous $k$ circuits \\
         $P_k$ & Uncertainty estimate given $k$ previous observations \\
         $H_k[\hat{x}_{k-1}]$ & Jacobian of $h_k$ with respect to $x$ at the prior estimate $\hat{x}_{k-1}$ \\
         $v_k$ & Observation noise \\ 
         $R_k$ & covariance of the observation noise \\
    \end{tabular}
    \caption{Summary of important notation used in this work}
    \label{tab:notation}
\end{table}

\subsection{Gate set tomography}

Gate-model quantum computers implement quantum programs by executing quantum circuits. While these circuits are intended to comprise sequences of perfect logic operations, real-world quantum processors will inevitably suffer errors that degrade performance and distort the distribution of measurement outcomes. Gate set tomography is an experimental protocol and data analysis procedure designed to learn a self-consistent model of the full set of logic operations that can then be used to predict these distorted distributions for arbitrary circuits. 

In this work, we assume that the gate set consists of: one $n$-qubit state preparation, one $n$-qubit measurement with $N_E=2^n$ possible outcomes, and some number $N_G$ of distinct $n$-qubit quantum gates. The standard model of errors fit by GST assigns to each of these operations a mathematical object:
\begin{enumerate}
    \item State preparation: $\sket{\rho} \in \mathcal{B}(\mathcal{H})$, a $4^n$-dimensional (column) vector that is a vectorized density matrix, 
    \item Logic gates: $\left\{G_i: \mathcal{B}(\mathcal{H}) \rightarrow \mathcal{B}(\mathcal{H})\right\}_{i=1}^{N_g}$, each a $4^n\times 4^n$-dimensional \textit{process matrix},
    \item Measurement: $\left\{ \sbra{E_j} \in \mathcal{B}(\mathcal{H})^*\right\}_{j=1}^{N_E}$, each a $4^n$-dimensional dual (row) vector that is a vectorized measurement effect.
\end{enumerate}
Here $\mathcal{B}(\mathcal{H})$ is the space of bounded operators acting on the $2^n$ dimensional Hilbert space $\mathcal{H}$ of pure $n$-qubit quantum states. GST learns the matrix elements of these objects either directly or via a parameterized error model $\mathcal{M}: x \mapsto \big\{\vert \rho \rangle\rangle, \{G_i\}, \{\langle\langle E_j \vert\}\big\}$ for some parameter vector $x$. 

A convenient and interpretable family of parameterized models for quantum gates is expressed in terms of \textit{error generators} \cite{Blume-Kohout2022-pv}. In these models, a noisy gate $G_i$ is written as the target operation followed by a small error effect 
\begin{equation}
    G_i = e^{\sum_{j} [x(G_i)]_j L_j } \tilde{G}_i,
\end{equation}
where $\tilde{G}_i$ is the ideal unitary action of the gate, $x(G_i)$ is a vector of (real) error rates for gate $G_i$, and $\{L_j\}$ is a basis for a Lie algebra of trace-preserving gate errors. Additional inequality constraints may be applied to enforce complete positivity. When $[x(g_i)]_j = 0$ for all $j$ and all $G_i$, then there are no errors in the device and the gate set is equal to the target unitary gate set. In well performing quantum computers, i.e. those with high gate fidelity and low non-Markovian effects, the gates generally well approximate their target unitaries, so real-world error rates (the components of $x$) are typically $\ll 1$. For the purposes of this work, we  collect the error rates for all gates into a single vector $x = \bigoplus_i x(G_i)$. Knowledge of $x$ completely describes the gate set model. 

GST probes these error rates by defining and repeatedly running a suitable set of quantum circuits. We define a depth-$d$ quantum circuit as an instruction to apply $d$ logical gates in sequence: $c^{(k)}=\left(c^{(k)}_d c^{(k)}_{d-1} \ldots c^{(k)}_2 c^{(k)}_1\right)$. The quantum process $C_k$ implemented by $c^{(k)}$ is modeled as the product of the $d$ process matrices corresponding to each layer of the circuit:  $C_k = G_{c^{(k)}_d} G_{c^{(k)}_{d-1}} \cdots G_{c^{(k)}_2} G_{c^{(k)}_1}$. The probability of observing outcome $j$ after running circuit $c^{(k)}$ is predicted to be:
\begin{equation}
    \text{Pr}(E_j|C_k) = \sbra{E_j} C_k \sket{\rho}. 
\end{equation}
In the language of the Kalman filter, this relationship defines the \textit{model observation function}. It is a map $h_k : \mathbb{R}^m \rightarrow \mathbb{R}^{2^n}$ from the vector of model parameters to the vector of modeled probabilities that is defined component-wise:
\begin{align}
    [h_k(x)]_j = \sbra{E_j} C_k \sket{\rho}
\end{align}

The particular class of circuits used by GST is designed to amplify all error parameters in a model. This is accomplished by choosing a list of \emph{fiducial} sequences of gates that rotate the native state preparation and measurement effects into an informationally complete set of effective state preparations and measurements. Additionally, we select a set of short gate sequences called \emph{germs} that, when repeated, collectively amplify all observable parameters of the error model. We construct circuits from these ingredients by sandwiching a repeated germ sequence between fiducial state preparation and measurement sequences. GST circuits thus take the form $ F^{\rm{meas}}_c G_b^p F^{\rm{prep}}_a $, where $F^{\rm{prep}}_a$ is a state preparation fiducial sequence, $F^{\rm{meas}}_c$ is a measurement fiducial sequence, and $G_b^p$ is a $p$-fold repeated germ sequence. Circuits are constructed for each germ and fiducial pair, and germ-powers are typically selected to be powers of two from 1 up to a maximum length dictated by both the desired estimation precision and quality of the logic operations (if the gates are good, we need long circuits to observe any errors). 

The $i$\textsuperscript{th} run of a circuit $c^{(k)}$ yields a single $n$-bit outcome string $s^{(k)}_i$ that is a categorical random variable sampled from the circuit's true outcome distribution. After running a particular circuit $N$ times, we compute the empirical distribution (the observed frequency of each $n$-bit string), which we denote as a $2^n$-dimensional vector $y_k$. We define $y_k$ component-wise:
\begin{align}
    [y_k]_j = \frac{1}{N} \sum_{i = 1}^N \mathbf{1}_j(s^{(k)}_i). 
\end{align}
Here $\mathbf{1}_j(s)$ is the indicator function. In the infinite shot limit, the observations $y_k$ converge almost surely \cite{Bertsekas2008-rl} to the true circuit outcome distributions. The goal of GST is to find a parameter estimate $\hat{x}_\text{GST}$ that brings all the model predictions $h_k(\hat{x}_\text{GST})$ as close to the observations $y_k$ as possible, typically as captured by the likelihood function:
\begin{align}
    \mathcal{L}(x \vert y_1, y_2, ..., y_k) &= \prod_{k=1}^K \text{Pr}(y_k| h_k(x))\\
        &\propto \prod_{k=1}^K \prod_{j=1}^N \sbra{E_j} C_k \sket{\rho}^{N [y_k]_j},
\end{align}
where the proportionality constant is a multinomial coefficient based on the count vector $N y_k$. 

\subsubsection{Gauge freedom}
A significant drawback of the error generator parameterization is that it is not unique---for any given model instance, there exist infinitely many equivalent models that all predict identical outcome probabilities for each circuit. Given one model parameterization, we may apply a gauge transformation that defined by an arbitrary invertible $4^n \times 4^n$ matrix $M$:
\begin{align*}
    \sket{\rho} &\mapsto M \sket{\rho}, \\
    G_i &\mapsto M G_i M^{-1}, \\
    \sbra{E_j} &\mapsto \sbra{E_j} M^{-1}.
\end{align*}
The predictions of the original model $\sbra{E_j} C_k \sket{\rho}$ are the same as the transformed model, even though the two models may appear completely different. While this gauge freedom does not impact the predictivity of the model, it does limit its \textit{interpretability} and \textit{observability}, as there are now extra ``gauge parameters'' in a model that do not correspond to any physically observable error process. 

The lack of observability of gauge parameters has significant impact on the performance of an online estimation algorithm. The role of observability in filtering theory was first introduced by Kalman \cite{Kalman1963-bh}, and in the context of linear, time invariant systems it is straightforward to derive a canonical observation form that decouples the observable parameters from the unobservable parameters. Given a canonical observation form, one simply estimates the observable parts and ignores the unobservable parts. However, for nonlinear systems, such as GST, the problem of observability is much more difficult \cite{Kou1973-ma}. We resolve the convergence issues cause by unobservable parameters by basing our filtering algorithms on first-order gauge-invariant (FOGI) models \cite{Nielsen2022-jx} that ensure the parameters are observable. 

FOGI models are constructed by considering small gauge transformations and separating out the gauge transformation's trivial null space from the non-trivial row space at the target model. A convenient sparse basis may then be found through various techniques. As discussed below, we have found that basing our estimation procedure on FOGI models seems to dramatically increase the robustness of our estimation algorithm and decrease the sensitivity of the filter to its initial point estimate. 

\subsection{Linear Kalman filters and their extension}
\label{subsec:KalmanFiltering}

As mentioned above, Bayesian inference is a natural technique to investigate for online estimation, but without simplifying assumptions, the utility of Bayesian approaches is hampered by significant computational burden. Kalman filters overcome this computational burden by assuming linear dynamics and Gaussian priors and noise distributions, which ensures all distributions used in the estimation algorithm can be treated analytically and require no sampling. The linear Kalman filter is optimal when a system is linear and the noise is Gaussian. In the case of gate set tomography, the model is non-linear and the observation noise is multinomial. However, we employ a series of approximations to cast the GST estimation problem in such a way that we may employ an extension of the Kalman filter that benefits from a significant computational speedup and retains excellent performance in simulation. 

The linear Kalman filter is widely used to estimate a hidden, possibly evolving state ${x}$ and noisy observations ${y}$ that are \textit{linear} functions of the state perturbed by Gaussian noise. The dynamics and observations are thus \emph{linear Gaussian models}. The system model may be cast as either continuous or discrete, and it can be adapted to include the effect of changes in control parameters. Because quantum operations are typically discrete entities, we focus here on the discrete time formulation. The most general linear Gaussian evolution of a state is
\begin{equation}
{x}_{k+1} = {F}_{k} \cdot {x}_k + {B}_k \cdot {u}_k + {w}_k 
\end{equation}
where the state transition matrix ${F}_k$ models known system dynamics, ${u}_k$ is a control input vector, ${B}_k$ models the effects of controls, and ${w}_k$ is a zero mean Gaussian random variable with covariance ${Q}_k$ that models stochastic noise in the evolution of the state. In the context of GST, the state will capture the error rates of the system, so this general form of the state transition function could be used to model drift, non-Markovianity, and changes in control inputs. However, in this work we consider only static noise models, so we can restrict the dynamic model and assume that ${x}_k$ is static in time, i.e., that ${F}_k$ is the identity and ${u}_k$ and ${w}_k$ are zero vectors. In the context of GST estimation, the assumption of static dynamics corresponds to an assumption that the device is Markovian and that we never change the control inputs. Future work will investigate scenarios with non-Markovian dynamics or changing controls, which would allow for dynamic calibration of drift in a quantum processor. Because we assume a static state, we may write ${x}$ without subscript to refer to the ``true'' state, i.e., throughout the rest of this work 
\[
    x_{k+1} = x_k \equiv x.
\]
While the assumption of static dynamics may appear very strong at first, this assumption is the usual one in quantum tomography, where we assume that one may prepare identical copies of the state without drift or changes in the controls used to prepare the states. 

A Kalman filter for static state estimation further assumes a linear Gaussian observation model of the form
\begin{equation}
{y}_{k} = {D}_k \cdot {x} + {v}_k
\end{equation}
where ${D}_k$ is a the observation model, or in the language of \cite{Gu2021-zx} the design matrix, that models the linear relationship between the state and the observation and ${v}_k$ is a zero mean Gaussian random variable with covariance ${R}_k$ that models stochastic noise in the observation. The subscript $k$ here indicates that at each time step we can choose from among the various types of observations that we may make of the hidden state ${x}$, each corresponding to a distinct GST circuit used to interrogate the system. 

The goal of Kalman filtering is to produce an estimate $\hat{x}_{k}$ of the hidden state ${x}$, as well as the uncertainty in the estimate ${P}_k$, given a iterative sequence of observations $\mathcal{Y}_k \equiv {y}_1, ..., {y}_k$. The uncertainty ${P}_k$ is quantified as a covariance matrix between the estimate and the true parameters 
\[
    {P}_k = \mathbb{E}\Big[ (\hat{{x}}_k - {x})(\hat{{x}}_k - {x}_k)^T \Big]
\]
where the expectation is conditioned on the sequence of previous observations $\mathcal{Y}_k$. 

Given a discrete time, linear Gaussian model, there are various ways to derive the Kalman filtering equations, e.g. from the perspective of minimum expected mean square error \cite{Kalman1960-jj}, jointly Gaussian random variables \cite{Meinhold1983-oj}, or simply by multiplying out a Gaussian prior and likelihood per Bayes rule. Because our formalism assumes partial observations of a static state ${x}$, the Kalman filter equations will reduce to a somewhat simpler form than usual. We defer writing their explicit form until we have introduced the extended Kalman filter, which is actually used in this work. 

The extended Kalman filter applies to systems governed by a nonlinear dynamical and/or observation model. Assuming that that the state is static, as above, then there is no time evolution in ${x}$ and the estimation algorithm is based only on partial, nonlinear observations of the state
\begin{equation}\label{equ:nonlinear_observation}
    {y}_k = {h}_k({x}) + {v}_k
\end{equation}
where ${h}_k$ is a nonlinear observation function that replaces the role of the linear design matrix ${D}_k$ in the linear Kalman filter and ${v}_k$ is the same observation noise as in the linear filter. 

In order to pass from a nonlinear observation to a linear form amendable to the assumptions of the Kalman filter, the extended Kalman filter linearizes the observation function ${h}_k$ \textit{about the prior estimate} $\hat{{x}}_{k-1}$. Linearization means calculating a design matrix ${H}_k[\hat{{x}}_{k-1}]$ that is the Jacobian of ${h}_k$ with respect to the parameter vector at the prior estimate
\[
    {H}_k[\hat{{x}}_{k-1}] \equiv \bigg[ \frac{\partial {h}_k}{\partial x} \bigg|_{\hat{{x}}_{k-1}}.
\]
Whenever we write a design matrix ${H}_k$ without an argument, it should be assumed that the linearization is taken at the prior estimate $\hat{{x}}_{k-1}$. 

Equipped with a notion of linearization, we may now write out the extended Kalman filter equations that form the backbone of our estimation algorithm. The Kalman filter updates a prior estimate and covariance $\hat{{x}}_{k-1}$ and ${P}_{k-1}$ into a posterior estimate and covariance $\hat{{x}}_{k}$ and ${P}_{k}$ according to 
\begin{eqnarray}\label{equ:extended_Kalman_update}
    \hat{{x}}_k = \hat{{x}}_{k-1} + {K}_k ({y}_k - {h}_k( \hat{{x}}_{k-1}) ) \\
    {P}_{k} = ({I} - {K}_k \cdot {H}_k) {P}_{k-1} \label{equ:extended_Kalman_covar_update}
\end{eqnarray}
where the Kalman gain ${K}_k$ is defined as 
\begin{equation}\label{equ:Kalman_gain}
    {K}_k = {P}_{k-1} {H}_k^T ( {H}_k {P}_{k-1} {H}_k^T + {R}_k)^{-1}, 
\end{equation}
and ${R}_k$ and ${H}_k$ are, as before, the observation noise covariance and the Jacobian of the observation function with respect to the model parameters at the prior estimate.

\section{Kalman filters and gate set tomography}
 \label{sec:kf_gst}

\begin{figure*}
\begin{minipage}[t]{\textwidth}
\begin{minipage}{0.7\textwidth}
    \includegraphics[width=0.95\textwidth]{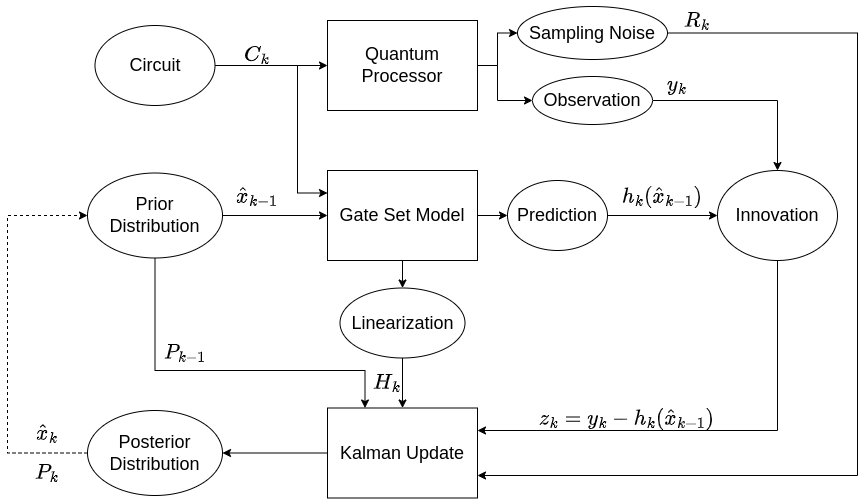}
    
\end{minipage}
\begin{minipage}{0.3\textwidth}\begin{algorithm}[H]
\caption{Kalman iteration}\label{alg:cap}
\textbf{Initialization:} \\
Gate set error model \\
Initial point estimate $\hat{x}_0$ \\
Initial uncertainty $P_0$ \\
\textbf{Output:} \\
Streaming estimate $\hat{x}_k$ \\
Streaming uncertainty $P_k$ \\
\textbf{Iteration:}
\begin{algorithmic}
\For{Each new circuit $C_k$} \\
    Take data $y_k$ \\
    $x_k \leftarrow x_{k-1} + K_k \cdot (y_k - h_k(\hat{x}_{k-1}))$ \\
    ${P}_{k} \leftarrow ({I} - {K}_k {H}_k) {P}_{k-1}$ 
\EndFor
\end{algorithmic}
\end{algorithm}
\end{minipage}

\end{minipage}
\caption{Kalman update algorithm structure and outline}
\label{fig:flowchart}
\end{figure*}

The Kalman filtering equations \ref{equ:extended_Kalman_update} and \ref{equ:extended_Kalman_covar_update} provide the backbone of our estimation routine, and Fig.~\ref{fig:flowchart} provides the detailed algorithmic structure of our method. In applying Kalman filtering to GST estimation, there are some key assumptions and approximations that we must make, which we address in this section. In particular, we discuss 1) the selection of an initial Gaussian prior, 2) the Gaussian approximation to the likelihood, and 3) the linearization of the observation model.

\subsection{Definition of initial priors}

A Kalman filter estimation routine requires initialization with an initial point estimate $\hat{{x}}_0$ that represents the initial guess and an initial covariance estimate ${P}_0$ that represents the estimated error in the guess. There are many choices for the initial point estimate, including starting at the target model, i.e. setting $\hat{{x}}_0 = {0}$, or seeding the filter with an estimate derived by linear regression or MLE on some smaller set of circuits (such as those of linear GST \cite{Nielsen2021-el}). One may also use a random Gaussian initial point centered about the target model with a predefined covariance. Other, more sophisticated techniques are also possible based on the outcome of randomized benchmarking data, such as the procedure used in \cite{Evans2022-jz}. In our numeric experiments, we found that estimating FOGI models is relatively robust to the choice of the initial point, and we are able to achieve good fits starting from the target model. 

The Kalman filter is relatively robust to over-estimation of the initial uncertainty, so there is some freedom in choosing the initial covariance estimate ${P}_0$. If ${P}_0$ is too small, then the filter may fail to converge, and if ${P}_0$ is too large, then the filter will explore more of parameter space early on in the estimation algorithm and thus converge more slowly. Ideally, the initial covariance should reflect the mean square error in the initial estimate. 

In our examples, we set ${P}_0$ to be equal to a scalar multiple of the identity matrix. We determined the magnitude of the covariance based on the outcome of a randomized benchmarking (RB) experiment, using the RB rate $r$ that estimates the average gate infidelity. Explicitly, we chose the initial covariance such that its trace is equal to $r$. In this fashion, we run a single RB experiment before we deploy streaming GST, which adds a constant overhead to the protocol. More sophisticated schemes to determine ${P}_0$ could likely be derived and is left for future work. 

\subsection{Gaussian likelihoods}

In the context of Kalman estimation for GST, the observations are the observed frequencies of circuit outcomes. In order to employ a Kalman filter, we must describe our observations as nonlinear functions of the state, perturbed by additive Gaussian noise as in Equation \ref{equ:nonlinear_observation}. To do so we appeal to the central limit theorem \cite{Bertsekas2008-rl}. In the limit of many circuit repetitions, $M$, the multinomial-distributed observations $y_k$ will be well approximated as a multivariate Gaussian random variable centered at the true circuit probability distribution $h_k(x)$ (the observation function evaluated at the true error parameters) with covariance
\begin{equation}\label{equ:true_covar}
    {y}_k \sim \mathcal{N}(h_k(x),R_k).
\end{equation}
where
\[
    R_k \equiv \frac{\text{diag}(h_k(x)) - h_k(x)h_k(x)^T}{M} 
\]

The fact that our observations are Gaussian distributed in the limit of many circuit repetitions means that we may approximate the likelihood function for a given circuit as: 
\[
    \text{Pr}(y_k | h_k(x)) \propto \exp\Big( -\frac{1}{2} (y_k - h_k(x))^T \cdot R_k^{-1} \cdot (y_k - h_k(x)) \Big).
\]
where $2^n$ is the size of the output space and we omit the standard multivariate Gaussian normalization constant. Thus observation likelihoods are approximately Gaussian when the number of samples taken is sufficiently large. The precise number of samples that must be taken will generally depend on the number of qubits and the dimension of the output space. 

In practice, we do not know a circuit's true probability distribution $h_k(x)$, so we require an estimate of the observation covariance in order to perform Kalman filtering in practice. The approach we take is to use the covariance of the conjugate Dirichlet distribution, as described in Ref. \cite{Audenaert2009-dr}. Given a multinomial distribution over $M$ trials with average sample vector ${y}$, then the conjugate Dirichlet distribution whose mode is equal to ${y}$ is uniquely defined as the Dirichlet distribution with pseudo-counts $\alpha$ equal to the observed count vector ${s}$ plus the vector of all ones ${1}$, i.e. $\alpha = {s} + {1}$, as in Laplace's rule of succession.  The resulting Dirichlet distribution will have covariance 
\begin{equation}\label{equ:Dirichlet_Covar}
    R_k \approx \frac{1}{M + d + 1}\bigg(\frac{\text{diag}(\alpha)}{M+d} - \frac{\alpha \alpha^T}{(M+d)^2}\bigg)
\end{equation}
where $M$ is the number of samples and $d$ is the dimension of the probability vector space. In this fashion, we match the covariance of our observations with the covariance of the conjugate Dirichlet distribution for the multinomial observation. It is important to note that this covariance is singular, which comes from the fact that the total counts is fixed, so we must use the pseudo-inverse in place of the usual matrix inverse. This singularity can pose some issues in filter design and we discuss methods to deal with the singularity of the Dirichlet covariance in Section \ref{sec:extensions}. 

\subsection{Linearization and circuit selection constraints}

\begin{figure}
    \centering
    \includegraphics[width=0.5\textwidth]{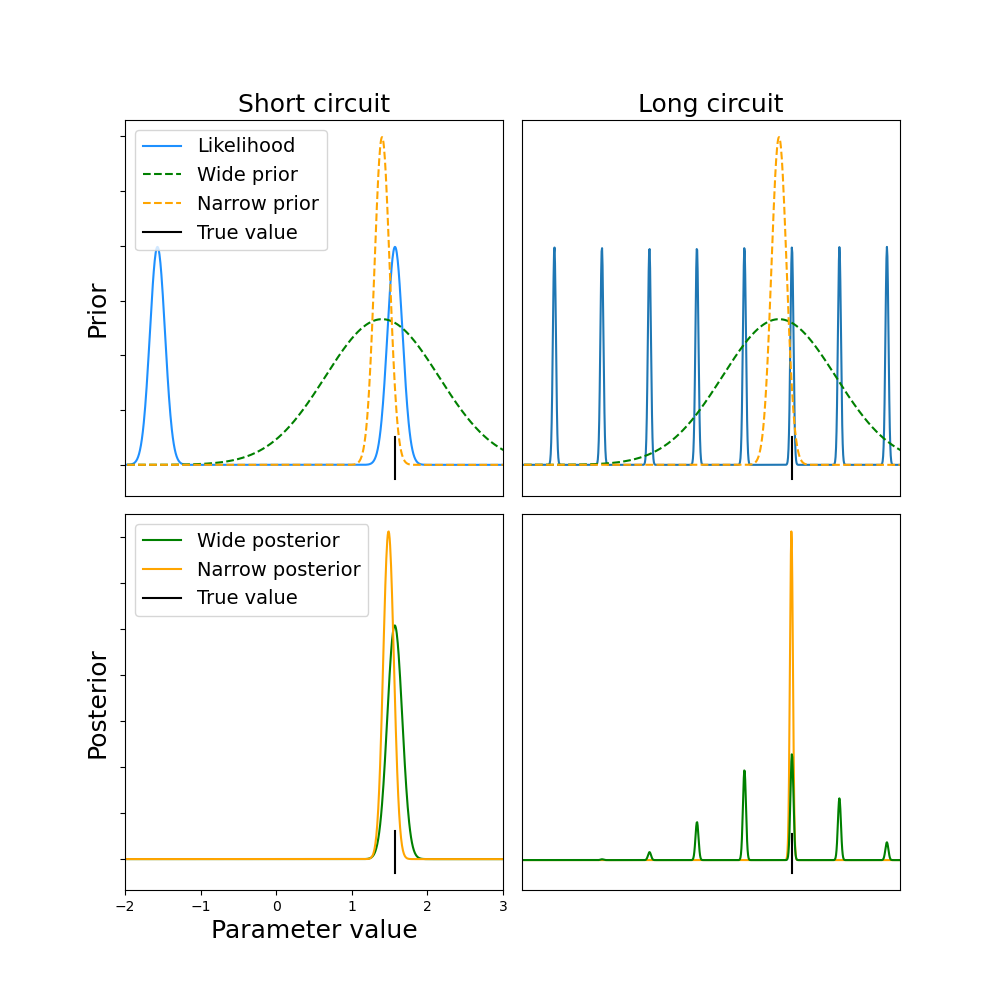}
    \caption{\textit{Influence of the prior distribution on Bayesian updates}. The top two plots show the prior and likelihood for a short and a long circuit for a simplified 1-parameter model, and the bottom two plots show the resulting posterior when calculated via Bayes rule. In the case of the short circuit, the wide circuit prior moves closer to the true value than the narrow circuit prior. In the case of the long circuit, the wide prior produces a multi-modal distribution when multiplied with the likelihood, which violates the assumptions of the Kalman filter, and the narrow prior results in a unimodal distribution that can be well approximated as a Gaussian. This example illustrates that, in order to assume Gaussian priors and Gaussian likelihoods needed for Kalman filtering, the length of the circuit should be selected such that the likelihood is unimodal on the principle support of the prior.}
    \label{fig:circuit_selection}
\end{figure}

Successful application of the extended Kalman filter requires accurately approximating the model observation $h_k$ with a linear expansion in the error parameters $x$. The technique of model linearization expands the observation function ${h}_k({x})$ about the prior estimate $\hat{{x}}_{k-1}$:
\begin{equation}\label{equ:prob_expansion}
    {h}_k({x}) = {h}_k(\hat{{x}}_{k-1}) + {H}_k[\hat{x}_{k-1}]\cdot ({x}-\hat{{x}}_{k-1}) + O(|{x}-\hat{{x}}_{k-1}|^2).
\end{equation}
In order for this approximation to hold, it must be the case that higher order variations in $h_k$ are negligible. However, the degree of nonlinearity in $h_k$ grows as a circuit's depth increases. Our heuristic for dealing with the increasing nonlinearity of $h_k$ is to start estimating with short circuits then feed in increasingly longer circuits as the expected estimate error shrinks. This way the observation function can be made to appear linear over the principle support of the prior, see Fig.~\ref{fig:circuit_selection}.  

Filtering on circuits in order from shortest to longest also addresses a particular kind of nonlinearity that arises in the GST circuit likelihoods. These likelihood functions can be approximately periodic in Hamiltonian error rates. This can violate the Gaussian assumption of Kalman filtering if the principle support of the prior (say the 95\% confidence region) spans more than a single period of the oscillation, see Fig.~\ref{fig:circuit_selection}. This issue also arises in robust phase estimation (RPE) \cite{Kimmel2015-kd} as longer circuits provide increased accuracy but only when one can use shorter circuits to identify the principle domain of the phase. By feeding in our circuits from shortest to longest, we ensure that the priors shrink at a rate comparable to the decrease in the period of the oscillation of longer circuit likelihoods. 


\section{Numerical results}

\label{sec:results}

To test and demonstrate the performance of extended filtering for online GST estimation, we have developed a Python class that interates with the \texttt{pyGSTi} \cite{Nielsen2020-sv} package to estimate gate set model parameters in a streaming fashion. In this section, we present numerical experiments that indicate that extended Kalman filtering is a promising candidate for real-time characterization of quantum processors. We estimate the parameters of tomographically complete 1-qubit and 2-qubit FOGI noise models using both iterative extended Kalman filtering and batched MLE. We find that the Kalman filter is able to achieve estimation accuracy that compares favorably with MLE on the two simulations presented here, as well as on numerous other simulations performed with different random noise models. We provide our code \cite{Marceaux2023-xs}, and invite the reader to test our methods on models of their choosing. 

The numerical experiments presented here follow the same basic steps summarized below: 
\begin{enumerate}
    \item A particular gate set is chosen. 
    \item A data generating model is chosen randomly.
    \item An GST experiment design is computed.  
    \item Experimental data is simulated using the data generating model. 
    \item The Kalman filter is applied to simulated observations from each circuit in turn.
\end{enumerate}

\begin{figure*}[h!]
\centering
\begin{subfigure}{0.45\textwidth}
\includegraphics[width=\linewidth]{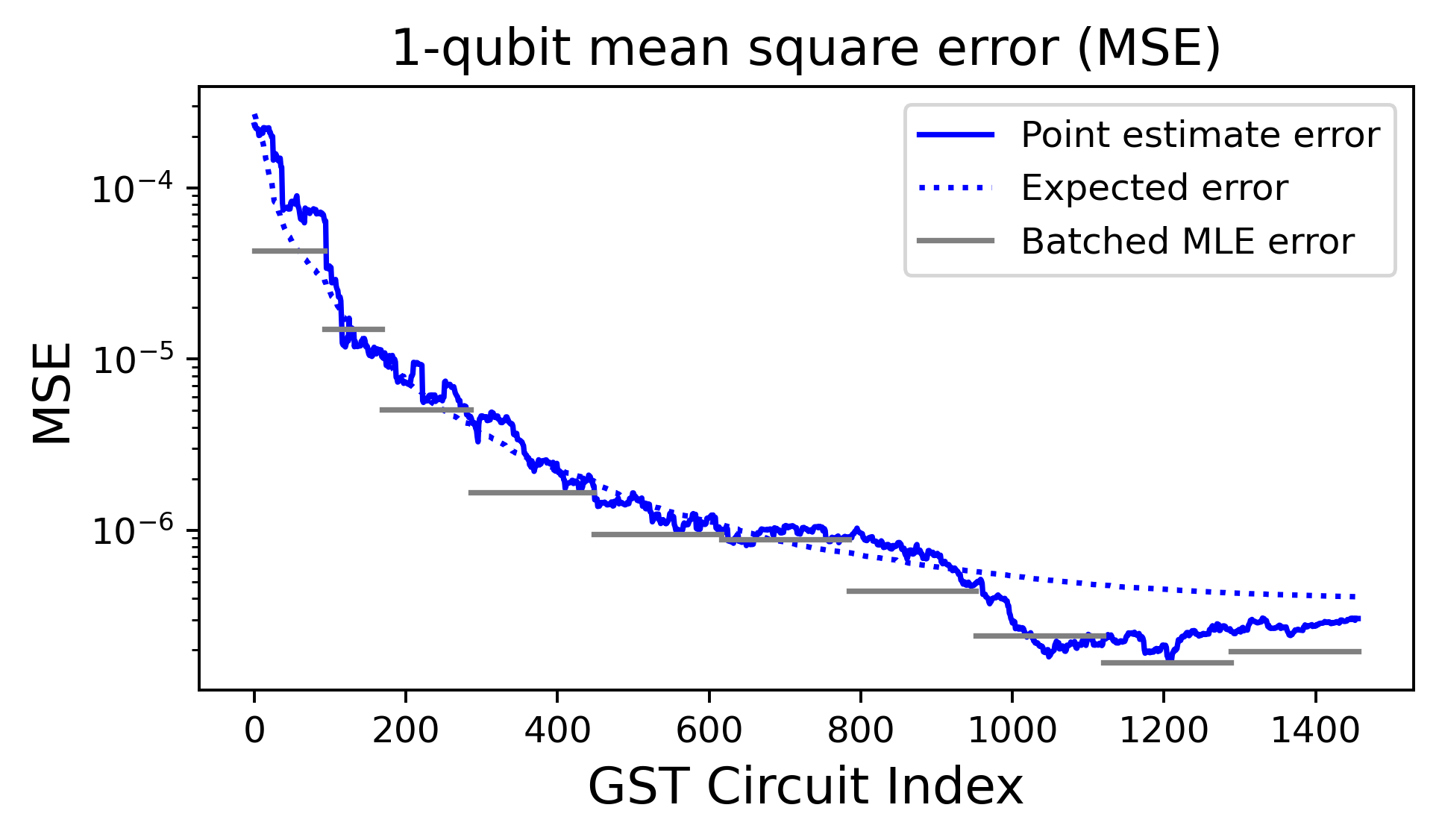}
\caption{}
\label{fig:1Qmse}
\end{subfigure}
\begin{subfigure}{0.45\textwidth}
\includegraphics[width=\linewidth]{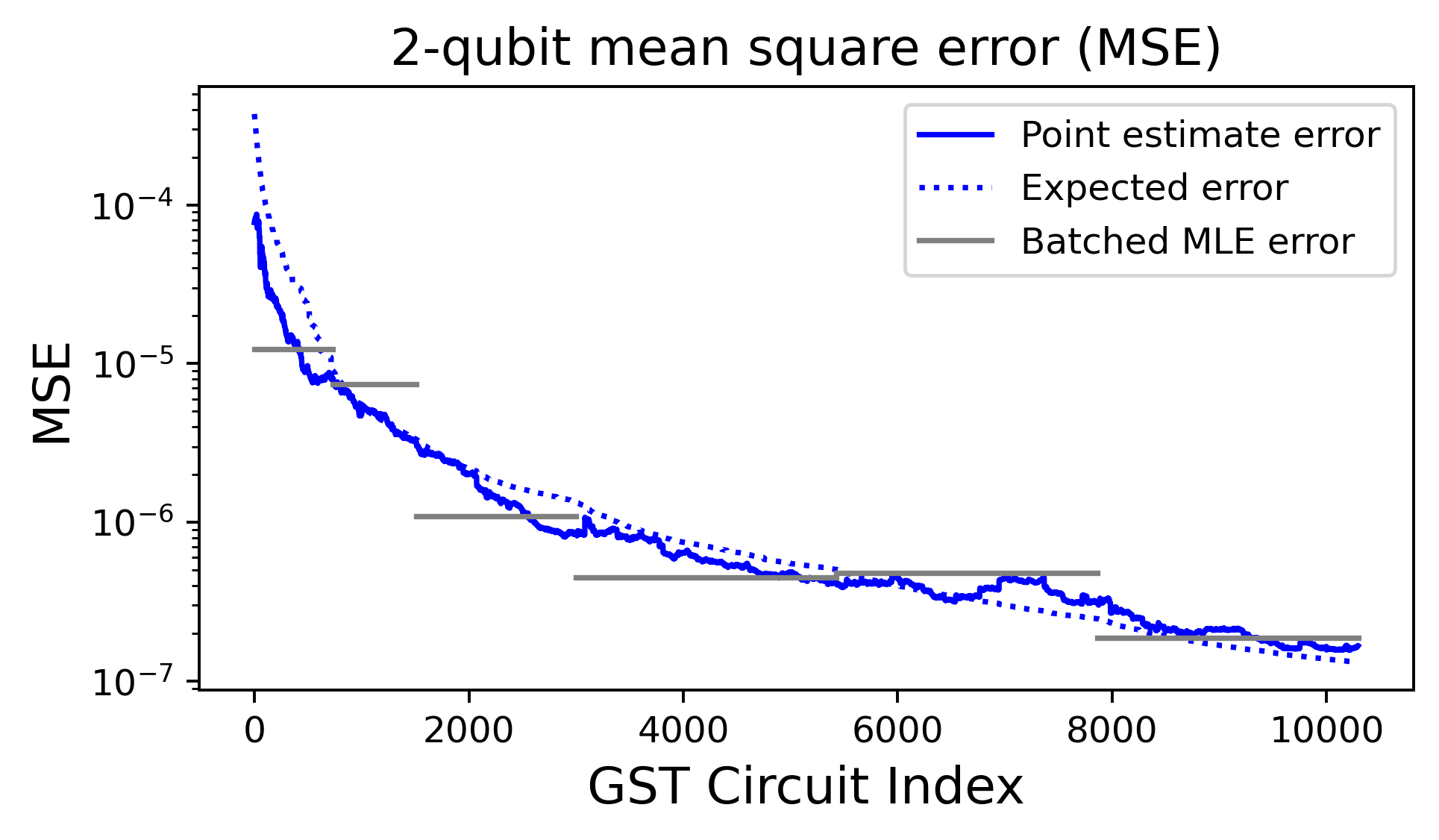}
\caption{}

\label{fig:2Qmse}
\end{subfigure}

\begin{subfigure}{0.45\textwidth}
\includegraphics[width=\linewidth]{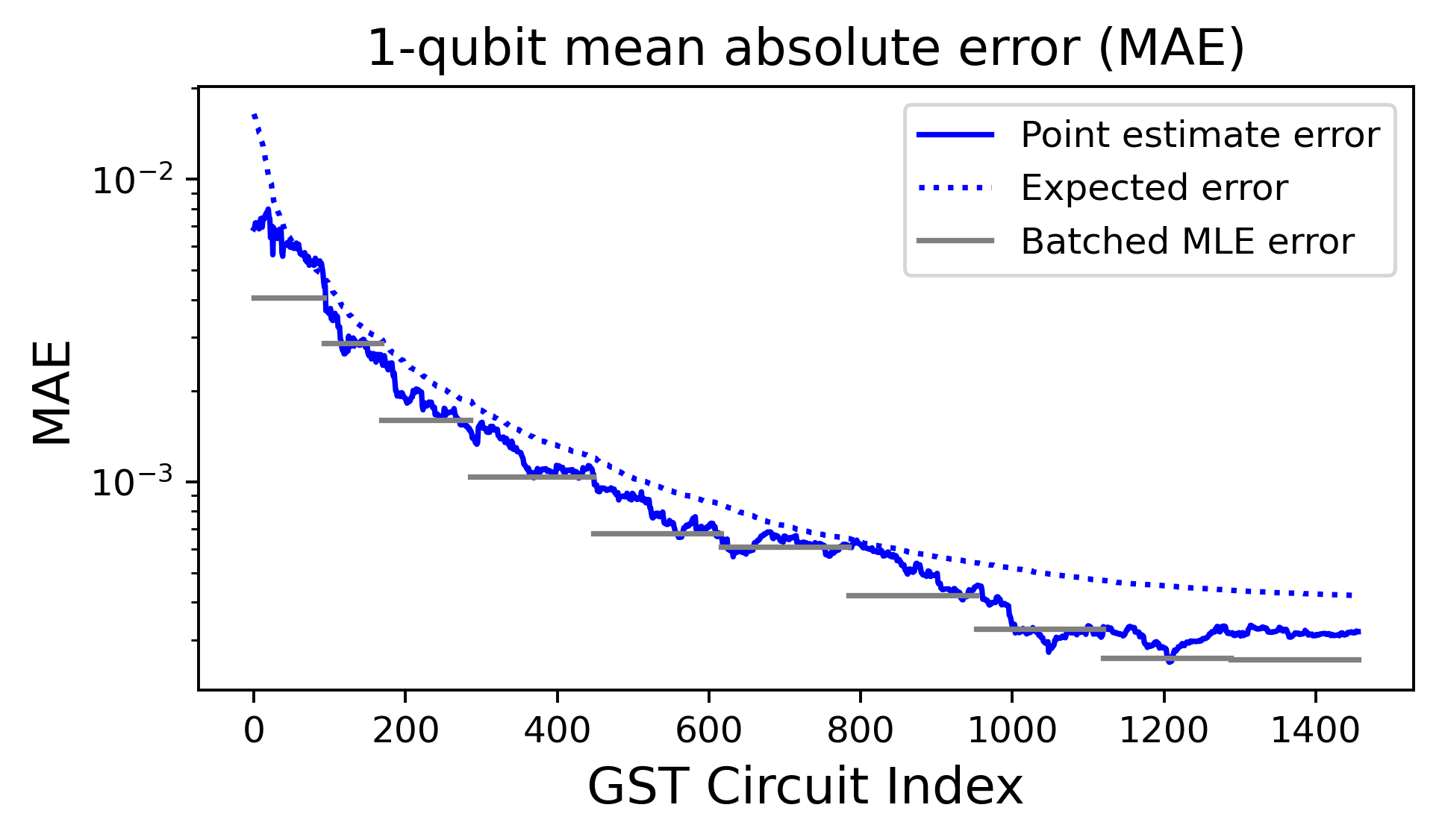}
\caption{}

\label{fig:1Qmae}
\end{subfigure}
\begin{subfigure}{0.45\textwidth}
\includegraphics[width=\linewidth]{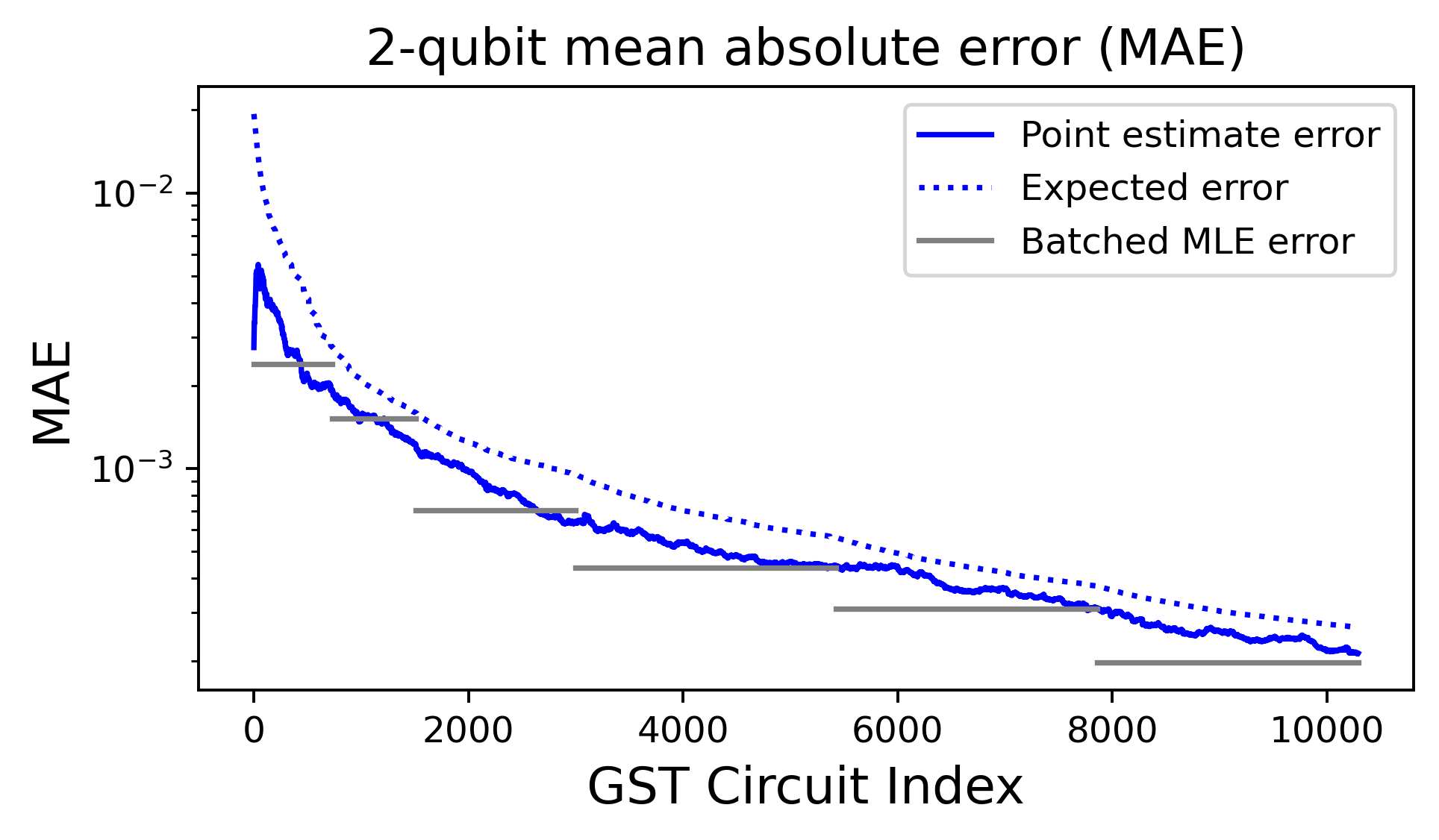}
\caption{}

\label{fig:2Qmae}
\end{subfigure}

\begin{subfigure}{0.45\textwidth}
\includegraphics[width=\linewidth]{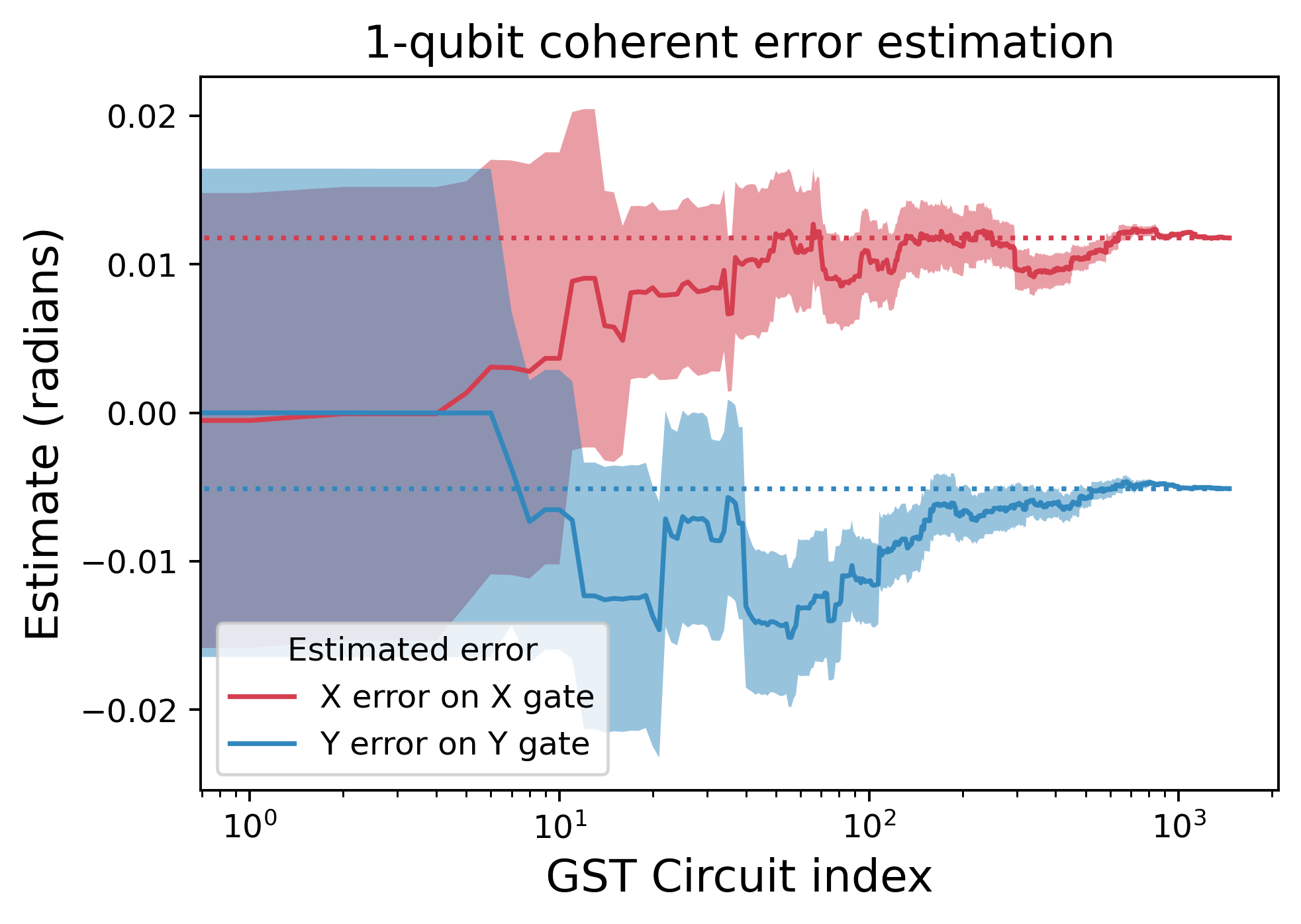}
\caption{}

\label{fig:1qoverrot}
\end{subfigure}
\begin{subfigure}{0.45\textwidth}
\includegraphics[width=\linewidth]{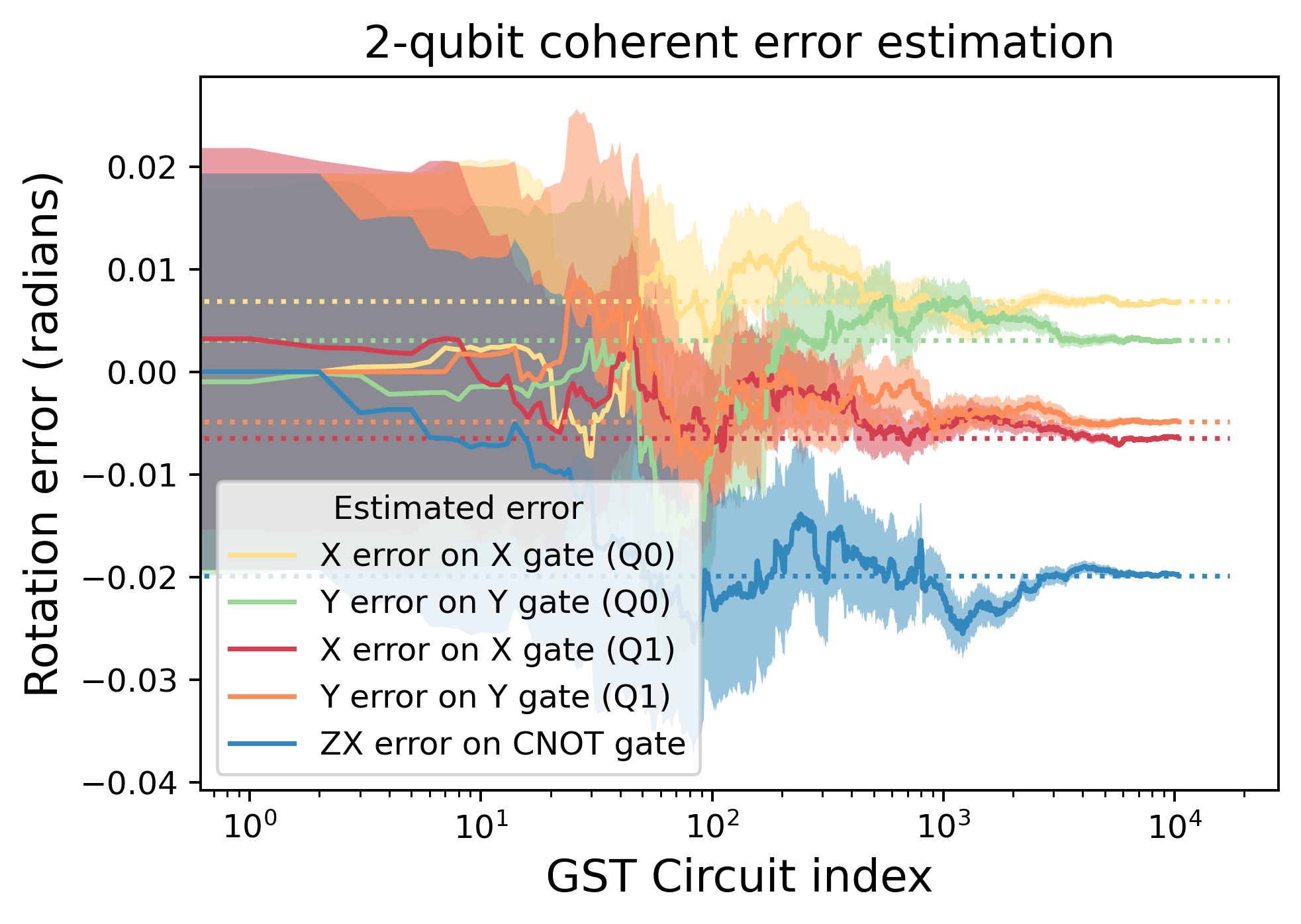}
\caption{}

\label{fig:2qoverrot}
\end{subfigure}

\caption{\textit{Numerical performance of streaming GST}. Plots (a)-(d) compare the convergence rates of the Kalman filter's point estimate with batched MLE point estimates under a metric of mean square error (MSE) and mean absolute error (MAE) between the point estimate and the parameters of the data generating model. The $x$-coordinates of the gray lines correspond to batches of germs of fixed power, and the batched MLE point estimates are calculated based on the observations from all data up-to and including the current batch. These plots also compare the evolution in the filters' MSE and MAE with their expected evolution given by $\text{Tr}(P_k)$ and $\text{Tr}(\sqrt{P_k})$ respectively. MSE is the natural metric for a Kalman filter since Kalman filters minimize the square of the expected error in the estimate, but MAE is a stronger performance metric that is more sensitive to small differences in parameters. Plots (e) and (f) display error in the estimate of particular Hamiltonian parameters that correspond to the types of errors we expect could be reduced with improved calibration. The dotted lines denote the ``true'' parameters that were used to generate the data. \label{fig:four}}
\end{figure*}

Our simulations are based on a 1-qubit gate set of $X$- and $Y$-rotations by $\pi/2$, and a 2-qubit gate set consisting of the same single-qubit gates and an additional controlled not (CNOT) gate from qubit 1 to 2. For convenience, we use a reduced H+S error model \cite{Blume-Kohout2022-pv} consisting solely of Hamiltonian and Pauli-stochastic errors, and reparameterize it using a FOGI representation. The restriction to an H+S model is not necessary, but simplifies our demonstration. We then generate a random data generating model with fixed Hamiltonian and stochastic error rates. We check that the model is completely positive and trace preserving (CPTP) and ensure that the average gate set infidelity is comparable to current devices. To highlight the ability of the Kalman filter to learn coherent errors, we also ensure that the coherent error rates contribute significantly to the infidelity. The choice of the initial covariance matrix was determined based on the outcome of a Clifford randomized benchmarking experiment \cite{Magesan2011-mz}, as described above.

Circuits were selected based on standard GST practice per the discussion in Sec. \ref{sec:background}. 
In practice, we found it useful feed in circuits from a batch of fixed germ power to the Kalman filter in a random order. GST experiment designs have inherent structure, such as many circuits for which the same germ is run with different fiducials, or that include only single-qubit gates. We found this structure caused distracting artifacts in the trajectory of the point estimate. However, we found no qualitative difference in the limiting behaviour of the estimate for randomized circuit batches and structured circuit batches. 

Figs. \ref{fig:1Qmse} and \ref{fig:2Qmse} display the evolution of the mean square error (MSE) in the filter model's mean point estimate as well as in the MLE point estimate, and Figs \ref{fig:1Qmae} and \ref{fig:2Qmae} display the mean absolute error. We also plot the expected MSE and the expected MAE in the estimate, which correspond to the trace of the covariance matrix $P_k$ and the trace of the square root of $P_k$, respectively. We find that the Kalman filter converges to the true model at a rate comparable to maximum likelihood estimation, and that the expected MSE and MAE evaluations are also consistent with the actual evolution. Our results indicate that Kalman filtering can achieve similar performance to batched MLE estimate.

To further illustrate the potential utility of our estimation algorithm for calibrations, we plot the evolution of a filter's estimate of specific Hamiltonian gate errors over time in Figs. \ref{fig:1qoverrot} and \ref{fig:2qoverrot}. In the case of single qubit gates, we plot on-axis over-rotation errors, e.g., for a $\pi/2$ rotation about $X$, the corresponding over-rotation error is an additional $\epsilon$ rotation about $X$. In the case of the CNOT gate, we plot the over-rotation of the $ZX$ Hamiltonian term. These results indicate that our filter is able to accurately estimate coherent, gate-specific errors, which are the types of errors that can be fixed with improved calibration and control. In particular, our technique also provides real-time uncertainty estimates of the errors, which will be useful in deriving real time calibration methods. 

An unexpected advantage of our approach is that that filter was able to process circuit outcome data at a relatively fast rate. We could process about 2-25 1-qubit circuits per second and 2-5 2-qubit circuits per second on a Dell xps laptop with an i5-8250U×8 processor. The processing rate goes down as the length of the circuit increases because of the increased complexity in recomputing Jacobians for longer circuits. However, we expect that the implementation efficiency could be significantly improved by exploiting the structure of GST circuits.

\section{Extensions and alternative approaches}\label{sec:extensions}

There are a number of alternative approaches and extensions to our core method that may be useful in future applications either in deploying on embedded hardware or in developing more refined filtering techniques, which we present here. In particular, we summarize techniques to (a) explicitly resolve the singularity of the Dirichlet covariance (rather than relying on a pseudo-inverse), (b) treat non-Markovian noise in the device, (c) speed up the estimation procedure with a significant increase in the required memory, and (d) increase the estimation precision when memory is limited. We have additionally investigated the sigma point (or unscented) Kalman filter \cite{Julier1997-mu}, and found that it achieves comparable performance to the extended Kalman filter. 

The Dirichlet covariance matrix that we use to model observation noise ${R}_k$ is singular. This singularity means that we must employ a more expensive pseudo-inverse in our estimation routine and that useful matrix factorizations, such as the Cholesky decomposition, cannot be applied. We see two potential amendments to our method so as to deal only with invertible matrices: 1) project out the singularity and consider only the invertible part of the covariance matrix, or 2) base our covariance estimate for the observation on a Poisson rather than Dirichlet distribution. Details of the first approach may be found in \cite{Audenaert2009-dr}, where the explicit form of the required projection operators is provided. The second approach is inspired by a Poisson Kalman filter that was derived to estimate disease transmission rates \cite{Ebeigbe2020-ss}. The Poisson Kalman filter replaces our definition of ${R}_k$ in Equation \ref{equ:Dirichlet_Covar} with the form
\begin{equation}\label{equ:Poisson_covar}
 {R}_k = \frac{\text{diag}(\alpha)}{(M+d)^2}
\end{equation}
where, again, the pseudo-counts vector $\alpha$ is the observed counts plus a vector of all 1's, $M$ is the number of samples, and $d$ is the dimension of the output space. As a diagonal positive matrix, the Poisson covariance estimate is clearly invertible. The key difference between the Dirichlet and the Poisson covariance estimates is that the Poisson covariance does not subtract the dyad $\alpha \alpha^T$, which is the cause of the singularity in the Dirichlet covariance. The Poisson form assumes that the samples in a batch are uncorrelated with one another, while the Dirichlet covariance includes information about correlations that arise due to the fixed shot count. We tested the Poisson form of the covariance in simulation and found little practical difference in the convergence rates of the point estimates of the two different filters, but more work is needed to fully understand the impact of this change.

In this work, we developed a Kalman filter for parameter estimation of \textit{static} GST parameters. However, error rates in real devices often display some amount of drift. To capture this drift, or even to provide robustness against more general non-Markovianity, we must relax the assumption of static dynamics. The Kalman filter conveniently has this ability already baked into its framework. Recall the ${Q}$ covariance matrix introduced in Section \ref{subsec:KalmanFiltering} models stochastic drift in the state. We previously assumed this covariance to be the all zeros matrix to reflect the fact that the gate set parameters were not changing over time. However, it would be straightforward to implement an extended estimation algorithm with a non-zero ${Q}$ matrix, which would explicitly allow for the possibility of Brownian parameter drift in the model assumptions. The exact form of the ${Q}$ matrix would naturally be specific to each particular system, and techniques for determining its form are left for future work. 

The most expensive step in our filtering routine is the calculation of observation function Jacobians, which must be reevaluated at the current estimate for each new circuit. The lattency of our algorithm may be significantly improved by approximating these first derivatives by a second-order Taylor expansion about some reference state. In this approach, the Jacobian at a point $\hat{{x}}$ may be approximated per
\begin{equation}
    {H}_k[\hat{{x}}_{k-1}] \approx {H}_k[\hat{x}_\text{ref}] + {A}_k[\hat{x}_\text{ref}] \cdot (\hat{{x}}_{k-1} - \hat{x}_\text{ref})
\end{equation}
where ${A}_k[\hat{x}_\text{ref}]$ is the Hessian or second order variations calculated at a reference point $\hat{x}_\text{ref}$. Instead of calculating ${H}_k[\hat{{x}}_{k-1}]$ every time we observe a circuit, we can precompute ${H}_k$ and ${A}_k$ at our chosen reference point and approximate the desired Jacobian with inexpensive matrix operations. One may envision a hybrid approach wherein a Kalman filter is run on batches of circuits and the model matrices are calculated for the next batch while the current batch is running, with a particular schedule that would naturally depend on the specifics of the system. Such a procedure would significantly the runtime latency of the estimation algorithm at the cost of increased prior computation and memory resources. If the available memory resources are not sufficient to store the Hessians, then one may consider using singular value compression on the Hessians, which, for GST circuits, will have a small subset of large singular values. 

The sigma point (or unscented) Kalman filter is a viable alternative to extended Kalman filtering that is particularly useful when model Jacobians are prohibitively expensive to calculate. We ran simulations using sigma point filtering and found that it achieved comparable estimation accuracy to extended Kalman filtering and with similar computational latency. In developing a sigma point Kalman filter, one may find difficulties employing the usual sigma point sampling algorithm, as the covariance of the Dirichlet distribution is singular. One may overcome these difficulties by either using a non-singular covariance for the observations, employing one of the techniques previously discussed in this section, or basing the sigma point sampling algorithm on the square root of the covariance instead of the Cholesky factor. 

In practice, we envision deploying our techniques on embedded hardware, where memory resources may be limited. When running a Kalman filter on real hardware such as an FPGA, it may be preferable to base an estimation protocol on the square root Kalman filter \cite{Grewal2010-pm}, which uses a Cholesky factor in place of the usual covariance matrix that appears in the estimation protocol. Because a Cholesky factor has a quadratically better condition number than a covariance matrix, the square root Kalman filter has a quadratically better estimation precision in the presence of limited memory. While some technical details of the estimation procedure change between the usual and the square root form of a Kalman filter, the higher level discussion and practical considerations discussed in this work should not change.




\section{Discussion}

\label{sec:discussion}

In this work, we developed an extended Kalman filter for quantum gate set tomography estimation. We demonstrated in simulation that our method can achieve similar estimation accuracy as the standard technique of batch maximum likelihood estimation. Our method additionally produces error bars for the estimate as a natural byproduct of the estimation procedure with no additional computation. Error bars in MLE analysis require expensive calculations based on Hessians of the likelihood function, which can take many hours or even days to calculate. We demonstrated that Kalman filtering based on first-order gauge invariant (FOGI) models without gauge degrees of freedom can reliably estimate model parameters even when seeded at the target model.

Adapting the extended Kalman filter to gate set tomography estimation required several key modifications to the GST experiment design, namely: (1) using a large number of samples per circuit to ensure approximately Gaussian observation noise, (2) ordering circuits by increasing depth to ensure that the model can be accurately linearized under the current uncertainty in the filter, and (3) using randomized benchmarking results to construct an initial Gaussian prior. These approximations also point to interesting future research questions including whether filters can be designed based on single shot circuit outcomes and investigating the validity of the linear approximation as a function of circuit length. 

Streaming gate set tomography based on the extended Kalman filter provides online model feedback, which is a key component in any closed-loop control framework. With our method, the user can use individual circuit outcome distributions to update an estimate the parameters of a gate set error model along with their uncertainty. Our protocol can be deployed on real devices and can process circuit outcomes at rates comparable with current circuit execution. 

This work represents a first step towards a unified closed-loop control algorithm for quantum processors. Towards this goal, our next steps include developing techniques for adaptive circuit selection that select the next circuit based on the current uncertainty in the filter, applying the filter in cases where the noise parameters change over time, and deriving control maps between changes in control parameters and changes in state parameters. These advances would then pave the way to a unified closed-loop control framework for arbitrary gate set quantum processors.


\section{Acknowledgements}

We thank Stefan Seritan, Corey Ostrove, and Erik Nielsen for helpful discussions and technical support. This material was funded in part by the U.S. Department of Energy, Office of Science, Office of Advanced Scientific Computing Research Early Career Research Program. JPM acknowledges additional support from National Science Foundation Award \#1747426 and the US Department of Energy, Office of Science, National Quantum Information Science Research Centers, Quantum Systems Accelerator. Sandia National Laboratories is a multimission laboratory managed and operated by National Technology and Engineering Solutions of Sandia, LLC, a wholly owned subsidiary of Honeywell International, Inc., for the U.S. Department of Energy's National Nuclear Security Administration under contract DE-NA0003525.

\printbibliography

@ARTICLE{Kimmel2015-kd,
  title     = "Robust calibration of a universal single-qubit gate set via
               robust phase estimation",
  author    = "Kimmel, Shelby and Low, Guang Hao and Yoder, Theodore J",
  journal   = "Phys. Rev. A",
  publisher = "American Physical Society",
  volume    =  92,
  number    =  6,
  pages     = "062315",
  month     =  dec,
  year      =  2015
}

@ARTICLE{Blume-Kohout2010-ax,
  title     = "Optimal, reliable estimation of quantum states",
  author    = "Blume-Kohout, Robin",
  journal   = "New J. Phys.",
  publisher = "IOP Publishing",
  volume    =  12,
  number    =  4,
  pages     = "043034",
  month     =  apr,
  year      =  2010,
  language  = "en"
}

@ARTICLE{Granade2016-gq,
  title     = "Practical {B}ayesian tomography",
  author    = "Granade, Christopher and Combes, Joshua and Cory, D G",
  journal   = "New J. Phys.",
  publisher = "IOP Publishing",
  volume    =  18,
  number    =  3,
  pages     = "033024",
  month     =  mar,
  year      =  2016,
  language  = "en"
}

@ARTICLE{Kalman1960-jj,
  title     = "A New Approach to Linear Filtering and Prediction Problems",
  author    = "Kalman, R E",
  journal   = "J. Basic Eng",
  publisher = "American Society of Mechanical Engineers Digital Collection",
  volume    =  82,
  number    =  1,
  pages     = "35--45",
  month     =  mar,
  year      =  1960
}

@ARTICLE{Audenaert2009-dr,
  title     = "Quantum tomographic reconstruction with error bars: a {K}alman
               filter approach",
  author    = "Audenaert, Koenraad M R and Scheel, Stefan",
  journal   = "New J. Phys.",
  publisher = "IOP Publishing",
  volume    =  11,
  number    =  2,
  pages     = "023028",
  month     =  feb,
  year      =  2009,
  language  = "en"
}

@ARTICLE{Magesan2011-mz,
  title    = "Scalable and robust randomized benchmarking of quantum processes",
  author   = "Magesan, Easwar and Gambetta, J M and Emerson, Joseph",
  journal  = "Phys. Rev. Lett.",
  volume   =  106,
  number   =  18,
  pages    = "180504",
  month    =  may,
  year     =  2011,
  language = "en"
}

@ARTICLE{Blume-Kohout2022-pv,
  title     = "A Taxonomy of Small {M}arkovian Errors",
  author    = "Blume-Kohout, Robin and da Silva, Marcus P and Nielsen, Erik and
               Proctor, Timothy and Rudinger, Kenneth and Sarovar, Mohan and
               Young, Kevin",
  journal   = "PRX Quantum",
  publisher = "American Physical Society",
  volume    =  3,
  number    =  2,
  pages     = "020335",
  month     =  may,
  year      =  2022
}

@ARTICLE{Evensen2003-lr,
  title    = "The Ensemble {K}alman Filter: theoretical formulation and
              practical implementation",
  author   = "Evensen, Geir",
  journal  = "Ocean Dyn.",
  volume   =  53,
  number   =  4,
  pages    = "343--367",
  month    =  nov,
  year     =  2003
}

@INPROCEEDINGS{Julier1997-mu,
  title      = "New extension of the {K}alman filter to nonlinear systems",
  booktitle  = "Signal Processing, Sensor Fusion, and Target Recognition {VI}",
  author     = "Julier, Simon J and Uhlmann, Jeffrey K",
  publisher  = "SPIE",
  volume     =  3068,
  pages      = "182--193",
  month      =  jul,
  year       =  1997,
  language   = "en",
  conference = "Signal Processing, Sensor Fusion, and Target Recognition VI"
}

@ARTICLE{Barrau2018-jo,
  title     = "Invariant {K}alman filtering",
  author    = "Barrau, Axel and Bonnabel, Silv{\`e}re",
  journal   = "Annu. Rev. Control Robot. Auton. Syst.",
  publisher = "Annual Reviews",
  volume    =  1,
  number    =  1,
  pages     = "237--257",
  month     =  may,
  year      =  2018,
  language  = "en"
}

@ARTICLE{Di_Matteo2020-in,
  title     = "Operational, gauge-free quantum tomography",
  author    = "Di Matteo, Olivia and Gamble, John and Granade, Chris and
               Rudinger, Kenneth and Wiebe, Nathan",
  journal   = "Quantum",
  publisher = "Verein zur Forderung des Open Access Publizierens in den
               Quantenwissenschaften",
  volume    =  4,
  number    =  364,
  pages     = "364",
  month     =  nov,
  year      =  2020,
  language  = "en"
}

@ARTICLE{Emzir2017-uo,
  title     = "A quantum extended {K}alman filter",
  author    = "Emzir, Muhammad F and Woolley, Matthew J and Petersen, Ian R",
  journal   = "J. Phys. A: Math. Theor.",
  publisher = "IOP Publishing",
  volume    =  50,
  number    =  22,
  pages     = "225301",
  month     =  may,
  year      =  2017,
  language  = "en"
}

@INPROCEEDINGS{Agarwal2019-jf,
  title     = "State Estimation of a Quantum System Using Extended {K}alman
               Filter",
  booktitle = "2019 International Conference on Cutting-edge Technologies in
               Engineering ({ICon-CuTE})",
  author    = "Agarwal, Nidhi and Sondhi, Akanksha and Singh, Ghanapriya",
  pages     = "97--100",
  month     =  nov,
  year      =  2019
}

@BOOK{Bertsekas2008-rl,
  title     = "Introduction to Probability",
  author    = "Bertsekas, Dimitri and Tsitsiklis, John N",
  publisher = "Athena Scientific",
  month     =  jul,
  year      =  2008,
  language  = "en"
}

@ARTICLE{Meinhold1983-oj,
  title     = "Understanding the {K}alman Filter",
  author    = "Meinhold, Richard J and Singpurwalla, Nozer D",
  journal   = "Am. Stat.",
  publisher = "Taylor \& Francis",
  volume    =  37,
  number    =  2,
  pages     = "123--127",
  month     =  may,
  year      =  1983
}

@ARTICLE{Nielsen2020-sv,
  title     = "Probing quantum processor performance with {pyGSTi}",
  author    = "Nielsen, Erik and Rudinger, Kenneth and Proctor, Timothy and
               Russo, Antonio and Young, Kevin and Blume-Kohout, Robin",
  journal   = "Quantum Sci. Technol.",
  publisher = "IOP Publishing",
  volume    =  5,
  number    =  4,
  pages     = "044002",
  month     =  jul,
  year      =  2020,
  language  = "en"
}

@ARTICLE{Gupta2021-kw,
  title     = "Adaptive filtering of projective quantum measurements using
               discrete stochastic methods",
  author    = "Gupta, Riddhi Swaroop and Biercuk, Michael J",
  journal   = "Phys. Rev. A",
  publisher = "American Physical Society",
  volume    =  104,
  number    =  1,
  pages     = "012412",
  month     =  jul,
  year      =  2021
}

@ARTICLE{Geremia2003-rn,
  title    = "Quantum {K}alman filtering and the {H}eisenberg limit in atomic
              magnetometry",
  author   = "Geremia, J M and Stockton, John K and Doherty, Andrew C and
              Mabuchi, Hideo",
  journal  = "Phys. Rev. Lett.",
  volume   =  91,
  number   =  25,
  pages    = "250801",
  month    =  dec,
  year     =  2003,
  language = "en"
}

@INPROCEEDINGS{Iida2010-eo,
  title     = "Robust quantum {K}alman filtering under the phase uncertainty of
               the probe-laser",
  booktitle = "2010 {IEEE} International Symposium on {Computer-Aided} Control
               System Design",
  author    = "Iida, Sanae and Ohki, Kentaro and Yamamoto, Naoki",
  pages     = "749--754",
  month     =  sep,
  year      =  2010
}

@BOOK{The_Analytic_Sciences_Corporation1974-zy,
  title     = "Applied Optimal Estimation",
  author    = "{The Analytic Sciences Corporation}",
  publisher = "MIT Press",
  month     =  may,
  year      =  1974,
  language  = "en"
}

@BOOK{Grewal2014-gs,
  title     = "{K}alman Filtering: Theory and Practice with {MATLAB}",
  author    = "Grewal, Mohinder S and Andrews, Angus P",
  publisher = "John Wiley \& Sons",
  month     =  dec,
  year      =  2014,
  language  = "en"
}

@ARTICLE{Grewal2010-pm,
  title    = "Applications of {K}alman Filtering in Aerospace 1960 to the
              Present [Historical Perspectives]",
  author   = "Grewal, Mohinder S and Andrews, Angus P",
  journal  = "IEEE Control Syst. Mag.",
  volume   =  30,
  number   =  3,
  pages    = "69--78",
  month    =  jun,
  year     =  2010
}

@ARTICLE{Kalman1963-bh,
  title     = "Mathematical Description of Linear Dynamical Systems",
  author    = "Kalman, R E",
  journal   = "Journal of the Society for Industrial and Applied Mathematics
               Series A Control",
  publisher = "Society for Industrial and Applied Mathematics",
  volume    =  1,
  number    =  2,
  pages     = "152--192",
  month     =  jan,
  year      =  1963
}

@ARTICLE{Ebeigbe2020-ss,
  title     = "{P}oisson {K}alman filter for disease surveillance",
  author    = "Ebeigbe, Donald and Berry, Tyrus and Schiff, Steven J and Sauer,
               Timothy",
  journal   = "Phys. Rev. Res.",
  publisher = "American Physical Society",
  volume    =  2,
  number    =  4,
  pages     = "043028",
  month     =  oct,
  year      =  2020
}

@INCOLLECTION{Daum2021-wi,
  title     = "Extended {K}alman Filters",
  booktitle = "Encyclopedia of Systems and Control",
  author    = "Daum, Frederick E",
  editor    = "Baillieul, John and Samad, Tariq",
  pages     = "751--753",
  year      =  2021
}

@BOOK{Simon2006-dd,
  title     = "Optimal State Estimation: Kalman, {H} Infinity, and Nonlinear
               Approaches",
  author    = "Simon, Dan",
  publisher = "John Wiley \& Sons",
  month     =  jun,
  year      =  2006,
  language  = "en"
}

@ARTICLE{Kou1973-ma,
  title     = "Observability of nonlinear systems",
  author    = "Kou, Shauying R and Elliott, David L and Tarn, Tzyh Jong",
  journal   = "Infect. Control",
  publisher = "Elsevier BV",
  volume    =  22,
  number    =  1,
  pages     = "89--99",
  month     =  feb,
  year      =  1973,
  copyright = "https://www.elsevier.com/open-access/userlicense/1.0/",
  language  = "en"
}

@INPROCEEDINGS{Nielsen2022-jx,
  title    = "First-order gauge-invariant error rates in quantum processors",
  author   = "Nielsen, Erik and Young, Kevin and Blume-Kohout, Robin",
  volume   =  2022,
  pages    = "M38.009",
  month    =  jan,
  year     =  2022
}

@ARTICLE{Gu2021-zx,
  title     = "Randomized Linear {Gate-Set} Tomography",
  author    = "Gu, Yanwu and Mishra, Rajesh and Englert, Berthold-Georg and Ng,
               Hui Khoon",
  journal   = "PRX Quantum",
  publisher = "American Physical Society",
  volume    =  2,
  number    =  3,
  pages     = "030328",
  month     =  aug,
  year      =  2021
}

@ARTICLE{Evans2022-jz,
  title     = "Fast {B}ayesian Tomography of a {Two-Qubit} Gate Set in Silicon",
  author    = "Evans, T J and Huang, W and Yoneda, J and Harper, R and Tanttu,
               T and Chan, K W and Hudson, F E and Itoh, K M and Saraiva, A and
               Yang, C H and Dzurak, A S and Bartlett, S D",
  journal   = "Phys. Rev. Appl.",
  publisher = "American Physical Society",
  volume    =  17,
  number    =  2,
  pages     = "024068",
  month     =  feb,
  year      =  2022
}

@ARTICLE{Nielsen2021-el,
  title     = "Gate set tomography",
  author    = "Nielsen, Erik and Gamble, John King and Rudinger, Kenneth and
               Scholten, Travis and Young, Kevin and Blume-Kohout, Robin",
  journal   = "Quantum",
  publisher = "Verein zur Forderung des Open Access Publizierens in den
               Quantenwissenschaften",
  volume    =  5,
  number    =  557,
  pages     = "557",
  month     =  oct,
  year      =  2021,
  language  = "en"
}

@MISC{Marceaux2023-xs,
  title  = "{Online gate set tomography with the extended Kalman filter}",
  author = "Marceaux, J P and Young, Kevin",
  month  =  jun,
  year   =  2023
}
\end{document}